# Superconducting transition temperature of $MgB_2H_{0.03}$ is higher than that of $MgB_2$

V.V. Flambaum [1], G.A. Stewart [2], G.J. Russell [1], J. Horvat [3], S.X. Dou [3]

1. School of Physics, University of New South Wales, Sydney NSW, 2052, Australia.
2. School of Physics, University College, UNSW, Australian Defence Force Academy, Canberra ACT, 2600, Australia.
3. Institute of Superconducting and Electronic Materials (ISEM), University of Wollongong, Northfields Avenue, Wollongong NSW, 2522, Australia.

**Abstract**
Hydogenation of $MgB_2$ powder has lead to an increase in the superconducting temperature, as determined by ac susceptibility. Applied dc fields reduced the transition temperature in the same ratio as for the pure powder.

Our present knowledge of the electronic and structural properties of the new binary intermetallic superconducting compound $MgB_2$, which has a critical transition temperature of 39K [1], has been comprehensively reviewed by Buzea and Yamashita [2]. The observation of the boron-isotope effect [3] on $T_c$, a simple BCS form for the energy gap from point-contact tunneling measurements [4-6], transport measurements [7-10], thermodynamic properties [7,11,12] and the phonon density of states [13.14] confirm that $MgB_2$ is most likely a phonon-mediated s-wave superconductor with intermediate [6,11.14] or strong coupling [12]. The high critical transition temperature being due to the high frequency of the optical boron-boron vibrational modes [15]. In fact, estimates of the phonon frequencies from electronic structure calculations [15-17] appear to be consistent with the high $T_c$ values, although An et al. [15] and Kong et al. [17] indicate that only a few modes have significant electron-phonon coupling.

Theoretically, the transition temperature of this type of superconductor may be increased if we increase the range of phonon frequencies, keeping the interaction strength approximately the same. This has been clearly seen for the boron-isotope effect [3]. In principle, the highest vibrational frequencies should appear for materials containing the lightest element, hydrogen. Therefore, the addition of hydrogen to $MgB_2$ should increase the available range of phonon frequencies which in turn may lead to an increase of the superconducting transition temperature. One may justify this assumption, for example, by using the existing data for the transition temperatures of palladium and palladium hydrides. Pd is a non-superconductor, while the transition temperatures for PdH and $PdH_2$ are 5K and 16K, respectively [18]. In the present paper we report a preliminary study of the transition temperature for hydrogenated $MgB_2$ powder.

Starting with commercial MgB$_2$ powder from Alfa Aesar, we hydrogenated 1g samples of the powder over a range of temperatures, pressures and times. The transition temperatures of the powdered samples before and after hydrogenation, as well as in a magnetic field of 0.5T, were determined from ac susceptibility measurements.

The hydrogen loading rig was essentially a stainless steel gas-handling system with Nupro valves, a small cylinder of ultra-high purity hydrogen, a reference volume of 0.5 litre, and a sample volume of 0.1 litre. The hydrogen pressure was monitored with a Barocel (0-10000 torr) piezoelectric pressure sensor mounted on a thermal base. The bulk of the sample volume is associated with a stainless steel finger of length 300 mm and outer diameter 19 mm that is connected to the rest of the system with a copper gasket seal. The MgB$_2$ powder was placed in a molybdenum boat that is pushed to the end of the finger. Once the finger was attached the system was evacuated and the reference volume filled to a predetermined pressure of hydrogen gas. This gas was then expanded into the sample volume and the end of the finger was heated in a small Heraeus split design tube furnace. After the sample had been heated to the appropriate temperature for the appropriate time, the furnace was switched off and the sample was allowed to cool. The difference in the initial and final hydrogen pressures at ambient temperature was used to estimate the amount of hydrogen that had been taken up by the MgB$_2$. However, given that only a very small amount of hydrogen appears to have been taken up, the accuracy of this estimate was influenced by relatively small changes in ambient temperature. Before the sample was removed, the finger was immersed in liquid nitrogen and then opened up to air. It is hoped that this process might "poison" the surface and inhibit loss of hydrogen once the specimen was removed from the hydrogen gas.

The ac susceptibility was determined using a conventional inductively coupled bridge system, with signal processing carried out automatically to give $\chi'$ and $\chi''$ separately. The powdered samples, ~ 40 mg, were contained in cylindrical teflon tubes sealed at one end with a small alumina block The tube was located on a copper peg in one half of the balanced secondary coil. The peg is coupled directly to the cold head of a closed cycle helium refrigerator. External dc magnetic fields could be applied either parallel or perpendicular to the ac exciting field. For the measurements of $\chi(T)$, the zero-field cooled sample was observed during warming at < 1K/min using a resistive heater built into the cold stage. Between each field sweep the samples were demagnetised by heating above the transition temperature.

For our initial hydrogenation study we selected temperatures of $600^0$C and $700^0$C, soak times of 15 and 30 minutes and pressures of 3 and 10 atmospheres. In all cases the transition temperature ($T_c$) of the MgB$_2$H$_x$ powder increased above that of the original powder. Figure 1 shows the increase of $T_c$ by ~ 0.5K for the material MgB$_2$H$_{0.03}$, which was hydrogenated at $600^0$C for 30 minutes at 3 atmospheres. Also shown in the figure is the effect of a dc magnetic field of value 0.5T applied perpendicular to the ac field. Increasing the ac field to 35.0G(rms) decreased Tc for both the pure and hydrogenated powders by the same relative value. The frequency dependence, over the range 200 - 20000Hz, for both powders showed that $T_c$ for the pure powder moved as expected but for the $MgB_2H_{0.03}$ powder $T_c$ did not move significantly.

The $MgB_2H_{0.03}$ powder was left under vacuum in the susceptometer for 14 days at ambient temperature with periodic cooling to 26K in order to perform a series of susceptibility measurements. The material was found to be quite stable over this time scale. The other hydrogenated powders studied were also found to be stable for at least 7 days, the normal measurement period.

Increasing the hydrogenation temperature to $700^0C$ and 10 atmospheres resulted in a smaller increase in $T_c$, ~ 0.4K and ~ 0.2K, for soak times of 15 and 30 minutes, respectively. Each of these materials had similar ac field and frequency dependences as for the $MgB_2H_{0.03}$ powder. Thus, lower temperature hydrogenation appears to result in larger Tc increase, but the optimum conditions have still to be determined.

These results again confirm the theoretical assumption of a $T_c$ dependence on high frequency modes for an electron-phonon mediated superconductor. The review article of Buzea and Yamashita [2] shows that all substitutionally doped compounds of $MgB_2$ give rise to a decrease in $T_c$, except maybe for zinc, which further indicates that hydrogenation has introduced phonon modes of higher frequency. Whether hydrogenation can increase $T_c$ substantially and the material, especially in the solid polycrystalline form, be stable over time, awaits further study.

Caption

Figure 1. AC susceptibility of pure $MgB_2$ (solid line) and hydrogenated $MgB_2H_{0.03}$ (dotted line) powder samples. Note the decrease in $T_c$ for each sample when a dc magnetic field of 0.5T is applied. The ac magnetic field is 0.35G, frequency 1kHz, the coils are slightly out of balance and the curves are displaced for clarity.

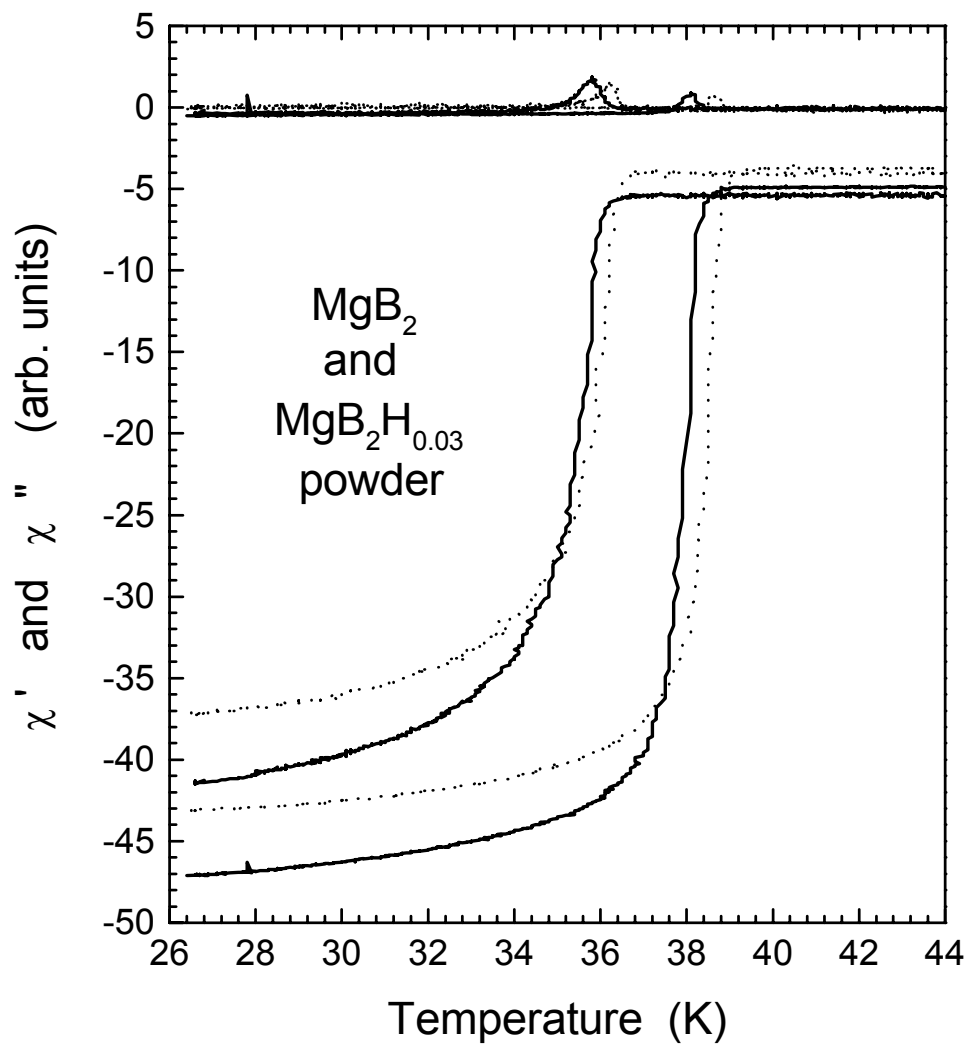